\documentclass[journal=macromol,manuscript=article]{achemso}

\usepackage{graphicx}
\usepackage{mathrsfs}
\usepackage{amsmath}
\usepackage{bm}
\usepackage{xcolor}
\usepackage{float}
\usepackage[marginal]{footmisc}
\usepackage{setspace}
\usepackage{amssymb}
\SectionNumbersOn

\newcommand{\be}{\begin{equation}}
\newcommand{\ee}{\end{equation}}

\title{Collapse of a single polymer chain:\\ Effects of chain stiffness and attraction range}
\author{Yanyan Zhu}
\affiliation{School of Physics and Astronomy, Tel Aviv University, Tel Aviv 69978, Israel}
\alsoaffiliation{Center for Physics and Chemistry of Living Systems, Tel Aviv University, Tel Aviv 69978, Israel}
\author{Haim Diamant}
\affiliation{School of Chemistry, Tel Aviv University, Tel Aviv 69978, Israel}
\alsoaffiliation{Center for Physics and Chemistry of Living Systems, Tel Aviv University, Tel Aviv 69978, Israel}
\author{David Andelman}
\affiliation{School of Physics and Astronomy, Tel Aviv University, Tel Aviv 69978, Israel}
\alsoaffiliation{Center for Physics and Chemistry of Living Systems, Tel Aviv University, Tel Aviv 69978, Israel}
\email{andelman@tauex.tau.ac.il}

\begin{document}
\begin{center}
\includegraphics[width=3.25in,height=1.75in,keepaspectratio]{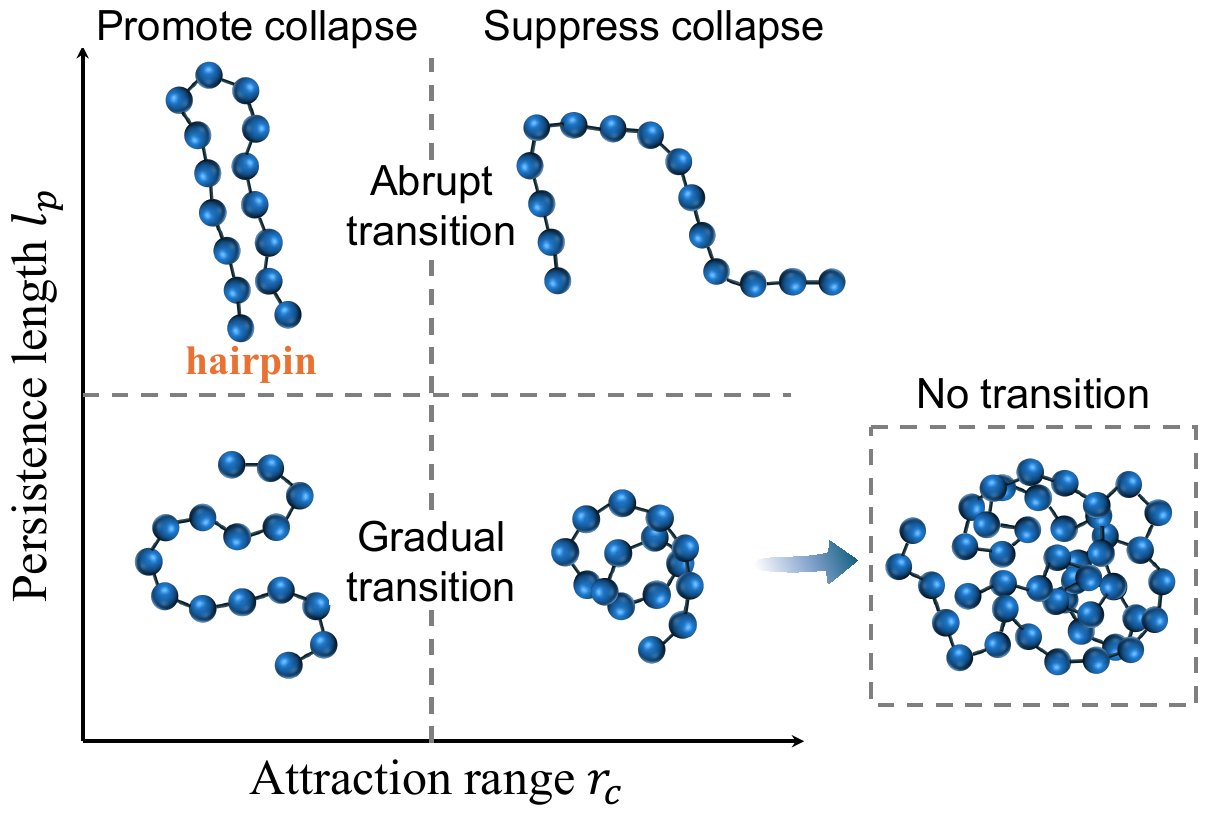}
\end{center}
\begin{abstract}
Chain-like macromolecules in solution, whether biological or synthetic, transform from an extended conformation to a compact one
when temperature or other system parameters change. This collapse transition is relevant in various phenomena, including DNA
condensation, protein folding, and the behavior of polymers in solution.  We investigate the interplay of chain stiffness and range
of attraction between monomers in the collapse of a single polymer chain. We use Monte Carlo simulations based on the pruned-enriched Rosenbluth method.
We demonstrate that the competition between the persistence length, $l_p$, and the range of attraction, $r_c$, 
determines whether the chain's collapse behavior resembles that of flexible chains or stiff ones. 
When $l_p$ is larger than $r_c$, the chain collapses sharply with decreasing temperature, whereas if $l_p$
is smaller than $r_c$, it contracts gradually. Notably, in the regime of small $l_p$ and large $r_c$,
this rounding into a gradual compaction persists upon increasing the chain length and may remain in place in the limit of infinite chain length.
Furthermore, for small $r_c$, the transition temperature ($\theta$-temperature)
increases with $l_p$, whereas for large $r_c$ the $\theta$-temperature decreases with $l_p$. Thus, stiffness promotes
collapse for small $r_c$ but suppresses it for large $r_c$. Our findings are in agreement with recent experiments on the contraction of
single-stranded RNA as compared to double-stranded DNA, and provide valuable insights for understanding polymer collapse 
and the essential polymer parameters affecting it.

\end{abstract}

\section{Introduction}
Single-chain collapse refers to the conformational transition of a
polymer chain from an extended coil-like structure to a compact
globular state in response to external changes such as temperature or
pH~\cite{Grosberg94,Rubinstein03,Kita05,Frerix06,Tanaka09}. 
This thoroughly studied phenomenon is fundamental in
polymer science and related to the commonly observed processes in
biomacromolecules, {\it e.g.}, protein folding and denaturation, as
well as in synthetic polymers~\cite{Nakata95,Sherman06,Xu06}.

The coil-to-globule transition was first predicted by Stockmayer in
1960~\cite{Stockmayer60} and later observed experimentally by Tanaka
and co-workers~\cite{Tanaka79,Tanaka80}.  Over the past
half-century, extensive research, 
encompassing both theoretical and experimental studies, 
has been conducted to investigate the collapse
of single chains. Lifshitz~\cite{Lifshitz68} and later Lifshitz,
Grosberg, and Khokhlov~\cite{Lifshitz78} pioneered the theoretical
framework for the coil-to-globule transition in homopolymers, which
was later refined by Grosberg and
co-workers~\cite{Grosberg92}. Subsequently, it was
validated through experiments on synthetic
macromolecules~\cite{Grosberg922,Wu98} using various experimental
techniques, including light scattering and
viscometry~\cite{Wu96,Baysal98,Ye07,Chakraborty18}.

While many uncharged synthetic polymers are flexible, polyelectrolytes and
important biopolymers, such as DNA, biofilaments, and certain
proteins, exhibit semiflexibility due to chain stiffness~\cite{Bustamante94,Gerrits21}. This behavior is
characterized by a persistence length $l_p$, which is the characteristic
distance over which the chain maintains its direction. The distinction
between flexible and semiflexible chains is very relevant to the chain
collapse, and a large body of research~\cite{Tatjana19,Arcangeli24} 
has focused on how the stiffness of semiflexible polymers affects their phase
behavior, including the formation of
different condensates such as rod-like and toroidal
structures~\cite{Leforestier09,Leforestier11,Seaton13,Marenz16,Majumder21}.

Theoretical studies have shown that the nature of the single-chain collapse transition
 depends sensitively on chain stiffness. For flexible polymers, the collapse corresponds to the $\theta$-transition, 
which is continuous (second-order) in the thermodynamic limit~\cite{Yang13,Wang17}. 
In contrast, for semiflexible or sufficiently stiff polymers, the transition can become discontinuous (first-order), 
reflecting a qualitatively different collapse mechanism~\cite{Post79,Grosberg94,Noguchi97,Rampf06}.

Regarding the effect of chain stiffness on the transition temperature
$T_\theta$, the natural expectation is that increased stiffness should
hinder compaction, leading to {\it lower} $T_\theta$. Indeed, this trend was
observed in many simulations~\cite{Seaton13,Marenz16,Majumder21}.  
However, an early mean-field theory by Doniach, Garel, and Orland~\cite{Doniach96}
predicted that $T_\theta$ should be {\it independent} of
stiffness.  Yet, a later work by Bastolla and
Grassberger~\cite{Bastolla97} suggested that $T_\theta$ should {\it
increase} with stiffness. These different results indicate that the
effect of chain stiffness on $T_\theta$ remains an intricate open question.

Much less work has been devoted to the effect of the range of
monomer-monomer attraction on the collapse transition. In a series of
Monte Carlo simulations, Binder, Paul, and co-workers~\cite{Binder08,Binder09,Binder13} 
investigated the phase diagram of a single polymer chain as a function 
of temperature and the range of a square-well attractive interaction. Their results
demonstrated that altering the interaction range could lead 
to a collapse transition into different conformational states, including a
collapsed globule and a compact crystallite.

The renewed interest in these fundamental single-chain issues has been recently 
motivated by experiments reported in ref~\citenum{Knobler23}, who investigated the condensation of long
double-stranded DNA (dsDNA), as compared to long single-stranded RNA
(ssRNA), in the presence of polyvalent cations. While the dsDNA
undergoes discontinuous condensation at a critical ion concentration,
the ssRNA exhibits gradual compaction over 
a wide concentration range. The large difference in stiffness between dsDNA and ssRNA probably underlies their distinct condensation behaviors. 
Thus, a general understanding of single-chain collapse scenarios would be valuable for tailoring polymer properties 
across diverse applications.

Motivated mainly by these recent findings~\cite{Knobler23}, 
we revisit the single-chain collapse by focusing on                   
the interplay between chain stiffness ($l_p$) and the range of intrachain attraction ($r_c$), 
which is found to determine the sharpness and temperature of chain compaction. 
While the distinction between continuous collapse in flexible chains and discontinuous collapse in stiff chains 
has been widely discussed in the literature, the role of the attraction range remains less examined. In this study, we demonstrate 
that the interplay between the persistence length, $l_p$, and the attraction range, $r_c$, determines whether the collapse resembles 
that of flexible chains or stiff chains, or results in a smoothing of the transition into gradual contraction. Therefore, the attraction range 
not only affects the collapse temperature but can also extend the critical temperature into a temperature range 
over which the chain contracts, with this behavior continuing as the chain length increases.

This paper is organized as follows. Section~2 describes the model and
simulation method. Section~3 presents our results and discussion, focusing on the transition sharpness and dependence of transition temperature on
polymer stiffness, attraction range, and chain length. Section~4 concludes with a
summary of the findings and their experimental implications. 
Finally, the Supporting Information presents a heuristic scaling argument~\cite{Haim01} 
for the rounding of the collapse transition due to the interplay between chain stiffness and condensing agents.

\section{Model}

\subsection{The pruned-enriched Rosenbluth method}
We model a polymer chain as a self-avoiding random walk (SAW) in three dimensions
(3D) on a simple cubic lattice.  Each monomer occupies one lattice
site, and two monomers cannot occupy the same site, mimicking self-avoidance.  To investigate
the collapse of a single chain, we employed the Pruned-Enriched
Rosenbluth Method (PERM)~\cite{Grassberger97}, 
which is an efficient Monte-Carlo (MC) algorithm to investigate single-chain collapse, as is explained next.
It is based on the Rosenbluth-Rosenbluth (RR) pioneering MC
algorithm~\cite{Rosenbluth55}, as well as on enrichment
techniques, in which sample attrition is reduced by replicating successful chains~\cite{Wall59}.

The chain configurations are built iteratively. 
At each step, the RR method adds a new monomer to the
partially built chain.  This $n$-th monomer occupies one of the
$m_n\le 5$ nonoccupied neighboring sites on the 3D cubic lattice.  
The normalized probability $p_k$ of selecting a specific site
$k\,{=}\,1,...,m_n$ for the $n$-th monomer is based on a Boltzmann weight
in the following way:
\be
  p_k={\rm e}^{-\beta E_k }/\sum_{l=1}^{m_n} {\rm e}^{-\beta E_{l} }\,,
\ee
where $E_k$ is the energy associated with the $k$ position and depends on the interaction with all nearby monomers,
and $\beta\,{=}\,1/k_{\rm B} T$ is the inverse thermal energy.
As the Boltzmann factor is included in the probability $p_k$,
the corrected weight $w_n$ associated with the $k$ position, is {\it independent} of $k$ 
\be
\label{eq:wei}
w_n = {\rm e}^{-\beta E_k}/p_k  = \sum_{l=1}^{m_n} {\rm e}^{-\beta E_{l} }\,,
\ee
and reflects the principle of `importance sampling',
where the Boltzmann bias in $p_k$ is compensated by the weight factor $w_n$ (see
Ref.~\citenum{Grassberger97} for more details). 

The total weight $W_N$ of a specific chain configuration composed of $N$ monomers
is given by
\be
\label{WN}
 W_N = \prod_{n=1}^N w_n\,. 
\ee

The RR method for simulating statistics of long chains is known to fail for long polymer chains due to two main
reasons~\cite{Grassberger97,Hsu11}. The first is the `attrition
problem', where steps with no valid placement cause most attempts to terminate, and consequently, 
only a small fraction of chains reach the full length, especially for large $N$. The second is due to large
fluctuations in the total  $W_N$ weight. The fluctuations in each of
the $w_n$ factors lead roughly to a log-normal distribution for
$W_N$~\cite{Grassberger97}, Eq. (\ref{WN}). It results in having too large weights for some chain configurations, 
while other configurations have too small weights. Consequently, the surviving chain configurations are
dominated by a single configuration, causing a substantial lack of sampling.

The {\it Pruned-Enriched} Rosenbluth Method (PERM) was developed to address
these issues~\cite{Grassberger97}.  The key idea behind PERM is that,
as the chain is grown, low-probability configurations are pruned, while
high-probability ones are enriched, depending on their $W_N$
weight. Two thresholds, $W_N^{+}$ and $W_N^{-}$, are introduced,
defining `high' and `low' weights, respectively. In his original work~\cite{Grassberger97}, 
Grassberger set the ratio of the enrichment and pruning thresholds to be approximately $10$. 
Such a choice of the ratio primarily affects the efficiency of PERM, 
but does not influence the simulation results themselves. 

The PERM simulations are then 
conducted in the following way. For $W_N >W_N^{+}$, the configuration is enriched by duplication,
so that two independent copies are propagated at the next growth step.
Each copy is assigned half the weight of its `parent', thereby ensuring
that the total statistical weight is conserved.  Conversely, for
$W_N<W_N^{-}$, the chosen configuration is conditionally `pruned' at
the next step in the following way. We choose a uniformly distributed random number $x \in
[0, 1]$. If $x < 1/2$, the configuration is discarded
(pruned). Otherwise, it is retained, and its weight is multiplied by a
factor of two, in order to preserve the total statistical weight.

\subsection{The attraction and bending energy}
In our model, monomers occupy distinct lattice sites, accounting 
for the excluded-volume repulsion. Therefore, the polymer chains represent a self-avoiding walk (SAW) on a 3D lattice. 
The energy $E_k$ in Eq.~(\ref{eq:wei}) is taken as a sum of two terms,
\be
   E_k = u_{\rm att}(r)+u_{\rm bend}(\theta)\,,
\label{eq:en}
\ee
where $u_{\rm att}(r)$ is the attractive energy between all pairs of monomers
separated by a distance $r$ in 3D space, and $u_{\rm bend}(\theta)$ is the bending
energy between two adjacent bonds of relative angle $\theta$.  

The attractive interaction, $u_{\rm att}$, is required to obtain the chain collapse transition.
As the form of this attraction does not matter substantially, we have used a simple form,
\be 
\label{uatt} 
u_{\rm att}(r) =
\begin{cases}
-{\varepsilon}\left(\frac{b}{r}\right), &\quad r \leq r_c \\
0, &\quad r > r_c\,.
\end{cases}
\ee
Here, $r_c$ is a cutoff distance defining the range of attraction,
$\varepsilon$ characterizes the attraction strength, and $b$ is the
bond length.

The bending-energy term is given by 
\be 
\label{ubend}
u_{\rm bend}(\theta)=\frac{\kappa}{b} \left(1-\cos \theta \right)\,,
\ee
where $\kappa$ is the bending stiffness, implying that the persistence
length is $l_p\,{=}\,\beta\kappa$. In our model, the angle $\theta$ 
can only take the discrete values $0,\pm \pi/2$ (having different weights),
associated with the 3D cubic lattice forward and sideways directions, 
while moving backward is not allowed for a SAW.

By substituting the expressions for $u_{\rm att}$ and $u_{\rm bend}$, 
Eqs. (\ref{uatt}) and (\ref{ubend}), into Eqs.~(\ref{eq:wei}) and (\ref{eq:en}), 
we obtain the weight $w_n$ of the $n$-th monomer,
\be
\label{wnk}
w_n= \sum_{l=1}^{m_n}\exp\left[ \frac{\beta\varepsilon b}{r_l} -\frac{l_p(1-\cos \theta_l)}{b} \right] \,.
\ee
Here $r_l$ is the position of the $l$-th candidate of the $n$-monomer, and $\theta_l$
 is the angle between the trial bond direction and the previous bond.

In the analysis below, we choose $l_p\,{=}\,\beta\kappa$ rather than $\kappa$ as the control parameter. As a
result, only the attraction term in the exponential depends on
temperature.  Hereafter, we will use
the rescaled (dimensionless) temperature $T^{*}\equiv 
k_{\rm B}T/\varepsilon\,{=}\,1/(\beta\varepsilon)$.

\section{Results and Discussion}

\subsection{Transition sharpness: Effect of $l_p$ and $r_c$}

\begin{figure*}[h!t]
{\includegraphics[width=1.0\textwidth,draft=false]{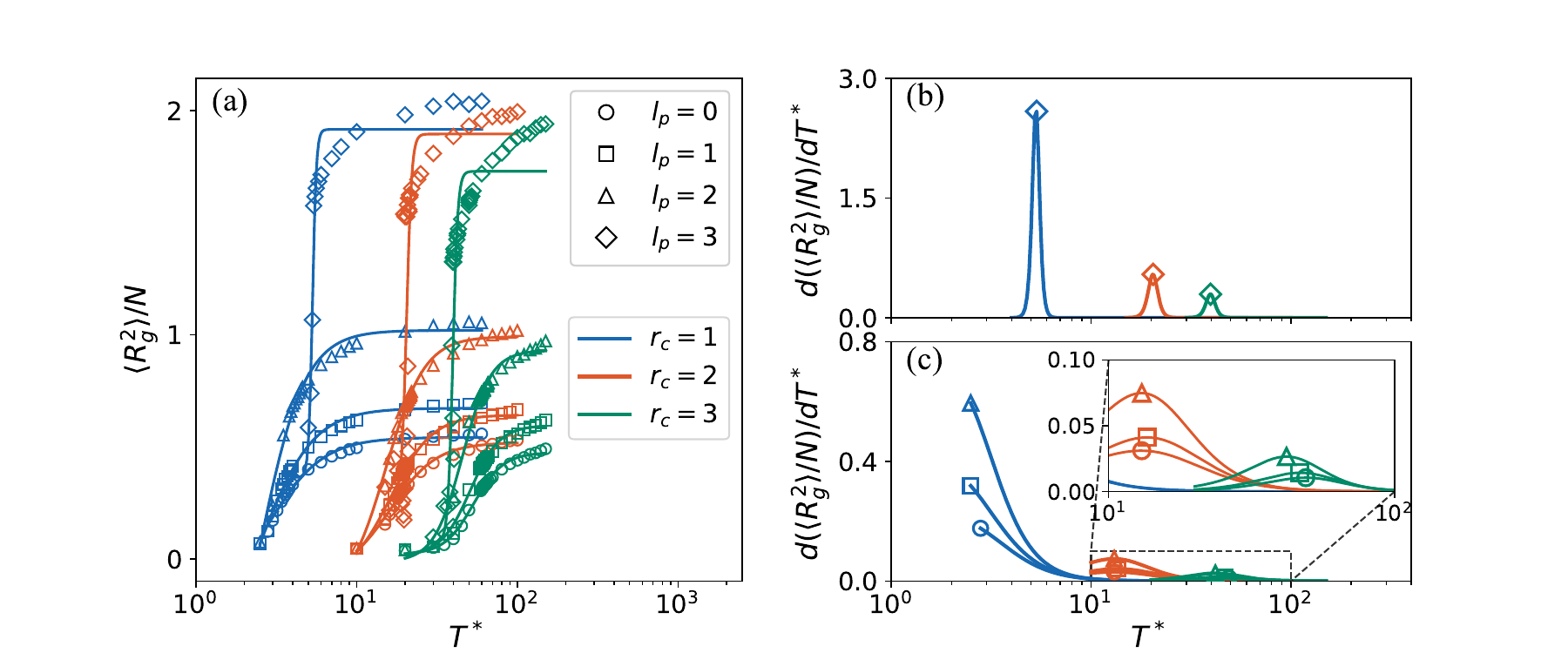}}
\caption{\textsf{ (a)Swelling factor $\langle R_g^2\rangle/N$ of the
    gyration radius as a function of rescaled temperature
    $T^{*}\equiv k_{\rm B} T/\varepsilon$, plotted on a semilog plot. Different
    symbols correspond to different values of the persistence length:
    $l_p\,{=}\,0$ (circle), $1$ (square), $2$ (triangle), and $3$
    (diamond), while different colors denote different values of
    the attraction range: $r_c\,{=}\,1$ (blue), $2$ (red), and $3$
    (green). The curves are fits to a
    sigmoid-like function [Eq.~(\ref{fit})]. For $l_p\,{=}\,3$, the fit is less accurate, but the derivative peak remains reliable.
 (b) and (c) Derivative $d(\langle R_g^2\rangle/N)/dT^*$ of the fitted curve as a function of the rescaled temperature $T^*$ 
    on a semilogarithmic scale, for different values of $l_p$ and $r_c$ (legend in (a)). (b) $l_p \,{=}\,3$; 
    (c) $l_p\,{=}\, 0$, $1$, $2$. The
    inset in (c) shows an enlarged view of the $r_c\,{=}\,2$ and $3$ curves over a narrow range of small derivative values, highlighting the very weak peaks.
    The data correspond to
    chains with $N\,{=}\,500$, and the bond
    length $b$ is taken as the unit length.  
    }}
\label{fig1}
\end{figure*}

We first conduct numerical simulations of polymer chains having a
length of $N\,{=}\,500$, with all lengths expressed in units of
the bond length $b$.  A total of approximately $10^5$ 
chains were generated using the PERM algorithm. Averages of the gyration radius squared, $\langle
R_g^2\rangle$, are computed over the ensemble of successful polymer
configurations. 
As the temperature decreases, the chain collapses as expected. We systematically
investigate the influence of chain stiffness $l_p$ and attraction
range $r_c$ on this chain collapse.

Figure~\ref{fig1}a shows the dependence of the swelling factor of the
gyration radius, $\langle R_g^2\rangle/N$, on the reduced temperature,
$T^*\equiv k_{\rm B} T/\varepsilon$, plotted on a semilog plot. 
The quantity $\langle R_g^2\rangle/N$ serves as a measure of the chain
average conformation, as will be further discussed in Sec.~\ref{subsec:results}. 
Lines of the same color (blue, red, green) represent
chains with the same attraction range $r_c$, while the same symbols
(circle, square, triangle, diamond) correspond to chains with the same
persistence length $l_p$. 

As seen in Fig.~\ref{fig1}a, the
transition changes from gradual to sharp as the persistence length
$l_p$ increases.  This trend is observed for chains with $r_c\,{=}\,1$
(blue), $r_c\,{=}\,2$ (red), and $r_c\,{=}\,3$ (green).
Conversely, for chains with the same
persistence length $l_p$ (denoted by the same symbol), 
the transition tends to become sharper as $r_c$ decreases.
The transition temperature exhibits different trends with increasing
$l_p$, depending on $r_c$, as is further discussed in Sec.~\ref{subsec:results}.

The sharpness of the transition clearly depends on the values of $l_p$ and $r_c$.
To quantify these differences, we fit in Fig.~\ref{fig1}a the swelling factor data as a function of
temperature to a sigmoid-like function,
\be
\label{fit}
\langle R_g^2\rangle/N =A + B \tanh \left[ C \ln (T^{*} /T_0) \right] \, , 
\ee
where $A$, $B$, $C$, and $T_0$ are fit parameters. 
Next, we calculate the derivative $d(\langle R_g^2\rangle/N)/dT^*$ (i.e. slope) of each fitted curve, as shown in Fig.~\ref{fig1}b and~\ref{fig1}c.
The maximum slope for each chain type, characterized by its $l_p$ and $r_c$ is indicated by symbols on each curve. 
For $l_p\,{=}\,3$ (diamond symbols in Fig.~\ref{fig1}a), although the global fit is less accurate, the location of the derivative peak remains robust. 
Figure~\ref{fig1}b corresponds to $l_p \,{=}\,3$, 
while Fig.~\ref{fig1}c shows results for $l_p \,{=}\, 0$, $1$, $2$. For $r_c\,{=}\,1$ in Fig.~\ref{fig1}c (blue curves), 
deviations from an ideal sigmoid occur at low $T^*$ due to the limited data range, but the derivative peak remains well defined.

It is evident that the maximum slope decreases with $r_c$ and increases with $l_p$. For small $l_p$ and large $r_c$ (see the inset of Fig.~\ref{fig1}c), 
the derivative $d(\langle R_g^2\rangle/N)/dT^*$ becomes very small, indicating that the sharp transition is effectively removed. 
This behavior arises because a larger $r_c$ leads to contraction of the entire polymer chain, thereby smoothing the transition 
and resulting in a more gradual compaction. The description of gradual contraction due to longer-ranged attraction is supported by a scaling argument, 
focusing on the case of chain collapse due to condensing agents, 
where $r_c$ is determined by the correlation length among the condensing agents \cite{Golestanian99,Haim01} (see Supporting Information). 
In the experimental systems, effective interactions are often influenced by solvent-mediated effects, condensing agents, or multivalent mobile ions, 
as seen in Ref.~\citenum{Knobler23}. For example, depletion interactions caused by condensing agents (e.g., PEG), 
as well as bridging interactions facilitated by multivalent counterions, amphiphilic molecules, or proteins~\cite{Bloomfield97, Gelbart00,Castelnovo04}, 
can produce effective attractions over an extended spatial range. These mechanisms can create an attraction range that surpasses the intrinsic persistence length. 
Therefore, the regime $r_c>l_p$ can be considered as an effective description of systems with such extended-range interactions.

\begin{figure*}[h!t]
{\includegraphics[width=1.0\textwidth,draft=false]{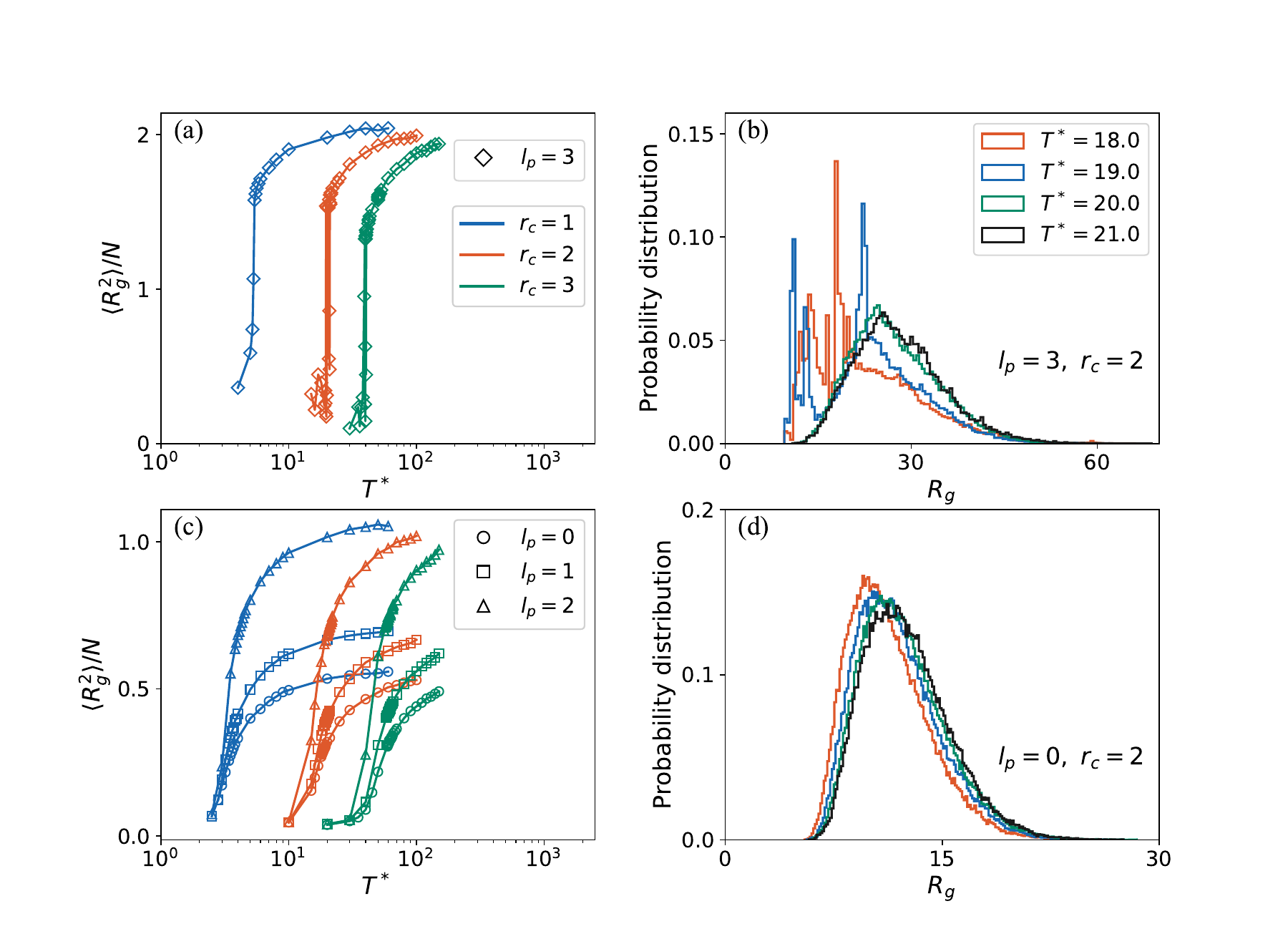}}
\caption{ \textsf{(a) and (c) Swelling factor $\langle R_g^2\rangle/N$
    as a function of the rescaled temperature $T^{*}\equiv k_{\rm B} T/\varepsilon$ on a semilog plot. (a) for $l_p\,{=}\,3$, and (c) for $l_p \,{=}\, 0$, $1$, $2$. 
    The three colors correspond to different $r_c$ values, as in Fig.~\ref{fig1}  (see also the legend). 
    Lines are guides to the eye. (b) and (d) Probability
    distribution histograms of $R_g$ for different temperatures:
    $T^{*}\,{=}\,18.0$ (red), $T^{*}\,{=}\,19.0$ (blue), $T^{*}\,{=}\,20.0$ (green),
    and $T^{*}\,{=}\,21.0$ (black). For all temperatures in (b) $l_p\,{=}\,3$ and
    $r_c\,{=}\,2$, and in (d) $l_p\,{=}\,0$ and $r_c\,{=}\,2$. Only the histogram outline is shown for clarity. Results
    are for chains of length $N\,{=}\,500$, and the bond length $b$ is taken as the unit length.
     }}
\label{fig2}
\end{figure*} 

For the stiffest chains in our study, $l_p\,{=}\,3$, we observe a sharp
transition as the temperature passes
through the transition. To further demonstrate the collapse behavior,
we replot in Fig.~\ref{fig2}a the data of Fig.~\ref{fig1}a for $\langle
R_g^2\rangle/N$ as a function of $T^{*}$ for the largest $l_p\,{=}\,3$ and $r_c\,{=}\,1,2$,
and $3$. 

In Fig.~\ref{fig2}b, we plot the entire probability distribution $P(R_g)$
for different $T^{*}$ values, where $l_p\,{=}\,3$ and $r_c\,{=}\,2$. As the rescaled temperature $T^{*}\,{=}\,k_{\rm B}
T/\varepsilon$ decreases from $21.0$ to $18.0$, the mean $R_g$ initially
decreases; yet, notably, at $T^{*}\,{=}\,19.0$, a bimodal distribution emerges
(blue histogram), indicating the coexistence of two distinct
phases. This observation suggests
that the collapse transition for stiff chains is first-order.
Our results are consistent with earlier PERM simulations by Grassberger~\cite{Bastolla97}, 
who showed that sufficiently large stiffness leads to a first-order coil–folded transition. 

For comparison, Fig.~\ref{fig2}c and \ref{fig2}d show the corresponding results for smaller persistence lengths ($l_p\,{=}\,0, 1,$ and $2$).
In Fig.~\ref{fig2}c, the swelling factor $\langle R_g^2\rangle/N$
 varies more smoothly with temperature, and the transition becomes progressively more gradual as $l_p$ decreases.
Consistently, the probability distributions in Fig.~\ref{fig2}d exhibit a single broad peak that shifts continuously with temperature, indicating a gradual compaction of the chain.
These observations are consistent with the experimentally observed discontinuous condensation of double-stranded 
DNA and continuous compaction of single-stranded RNA, supporting the relevance of our model to semiflexible polymer systems ({\it e.g.,} Ref.~\citenum{Knobler23}).

\subsection{Transition temperature: Effect of $l_p$ and $r_c$}\label{subsec:results}

\begin{figure*}[h!t]
{\includegraphics[width=0.85\textwidth,draft=false]{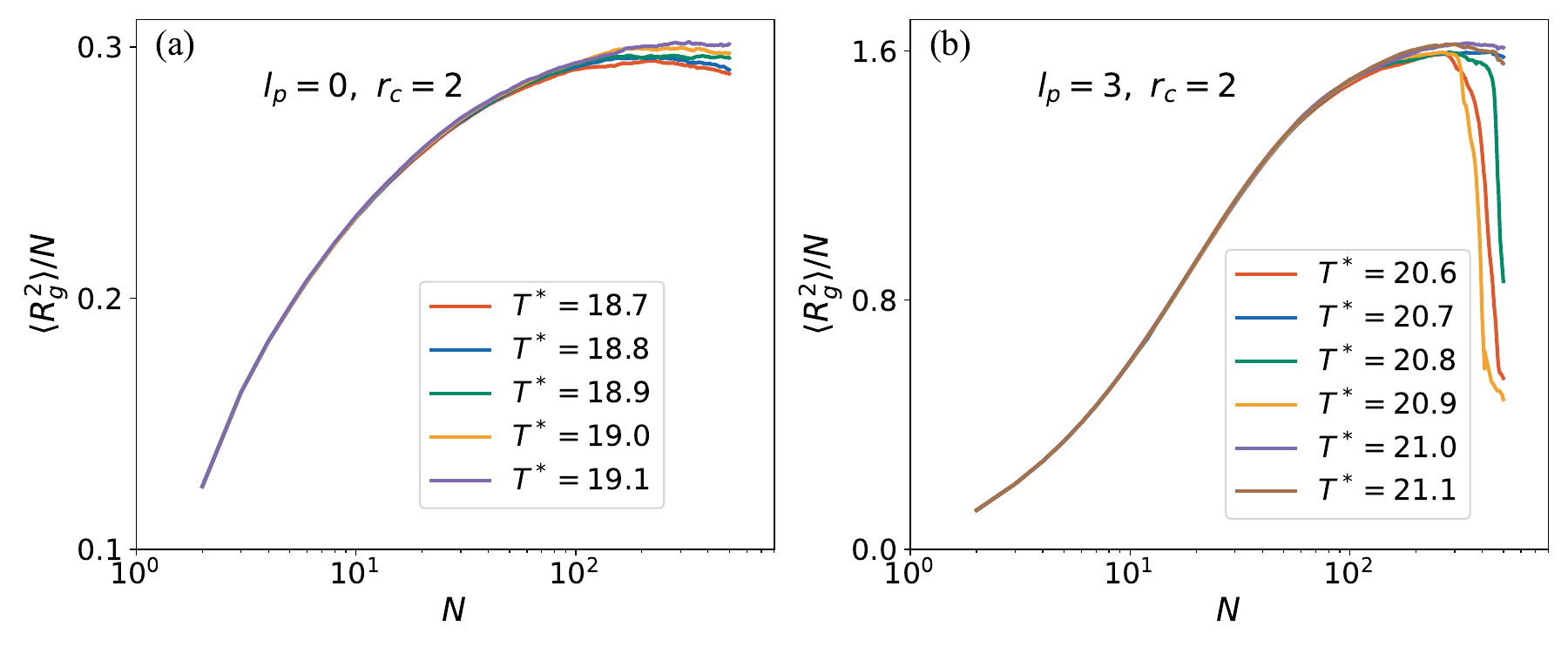}}
\caption{ \textsf{Swelling factor, $\langle R_g^2\rangle/N$, as a
    function of chain length, $N$, plotted on a semilog plot. (a)
    $l_p\,{=}\,0, r_c\,{=}\,2$ for $T^{*}\equiv k_{\rm B} T/\varepsilon\,{=}\,18.7,\, 18.8,\, 18.9,\, 19.0,$ and
    $19.1$ (see legend). The transition temperature is estimated as
    $T^{*}\simeq 18.9 \pm 0.1$. (b) $l_p\,{=}\,3, r_c\,{=}\,2$ for $T^{*}\,{=}\,20.6,\,
    20.7,\, 20.8,\, 20.9,\,  21.0,$ and $21.1$ (see legend).  The transition
    temperature is estimated as $T^{*}\simeq 21.0 \pm 0.1$. 
    }}
\label{fig3}
\end{figure*}

For $N\,{\gg}\, 1$,  $\langle R_g^2 \rangle$ scales with chain
length as $\langle R_g^2 \rangle \sim N^{2 \nu}$, where $\nu$ is the
Flory exponent.  Consequently, $\langle R_g^2\rangle /N \sim
N^{2\nu-1}$.  This quantity provides a sensitive probe of the chain
conformation. For $T>T_\theta$, the chain is swollen, and $\langle
R_g^2\rangle/N$ increases monotonously with $N$ ($\nu\,{>}\,1/2$). Whereas for
$T<T_\theta$, $\langle R_g^2\rangle/N$ initially increases with $N$ and then,
for large enough $N$, it decreases, indicating an asymptotic
value of $\nu\,{<}\,1/2$.  At $T{=}T_\theta$,  $\nu\, {\to}\,1/2$ for $N{\gg} 1$,
and $\langle R_g^2\rangle/N$ becomes approximately independent of $N$.
Therefore, we identify the temperature closest to $T_\theta$ as the
temperature where  $\langle R_g^2\rangle/N$ initially increases with $N$ and then
flattens out.

In Fig.~\ref{fig3}, the swelling factor $\langle
R_g^2\rangle/N$ is plotted as a function of $N$ 
on a semilog plot for different temperatures, while
keeping $l_p$ and $r_c$ fixed. Applying the above criterion, we identified the temperature at which the curve flattens out 
(the slope approaches zero) for large $N$ as the closest to $T_\theta$.
However, near the transition temperature, the curves for different
temperatures become similar, making visual determination difficult. Therefore, 
our criterion to identify $T^{*}_\theta$ is to choose the curve having an almost zero slope (taken here to be less than 0.01).
More specifically, in Fig.~\ref{fig3}a, the slopes for $l_p \,{=}\, 0$ and $r_c \,{=}\, 2$, at large $N$  and for $T^{*} \,{=}\,
18.8, 18.9$, and $19.0$ (blue, green, and orange lines, respectively)
are all below 0.01, indicating that $T^{*}_\theta \simeq 18.9 \pm 0.1$.

For stiffer chains, an illustrative example is shown in Fig.~\ref{fig3}b for $l_p\,{=}\,3$ and $r_c \,{=}\, 2$. 
Unlike the gradual collapse as a function of temperature, observed for flexible chains (Fig.~\ref{fig3}a), 
the stiffer chains exhibit a markedly different behavior 
due to a first-order abrupt transition. This is seen in the temperature range, $20.6 \le T^{*} \le 20.9$,
where the slope of $\langle R_g^2\rangle/N$ as a function of $N$ on a semilog plot
shows a pronounced drop at large $N$. However, $T^{*}_\theta$ is determined as in Fig.~\ref{fig3}a, 
as the temperature with the slope closest to zero, yielding $T^{*}_\theta \simeq 21.0\pm 0.1$ (purple line).

\begin{figure}[h!t]
{\includegraphics[width=0.45\textwidth,draft=false]{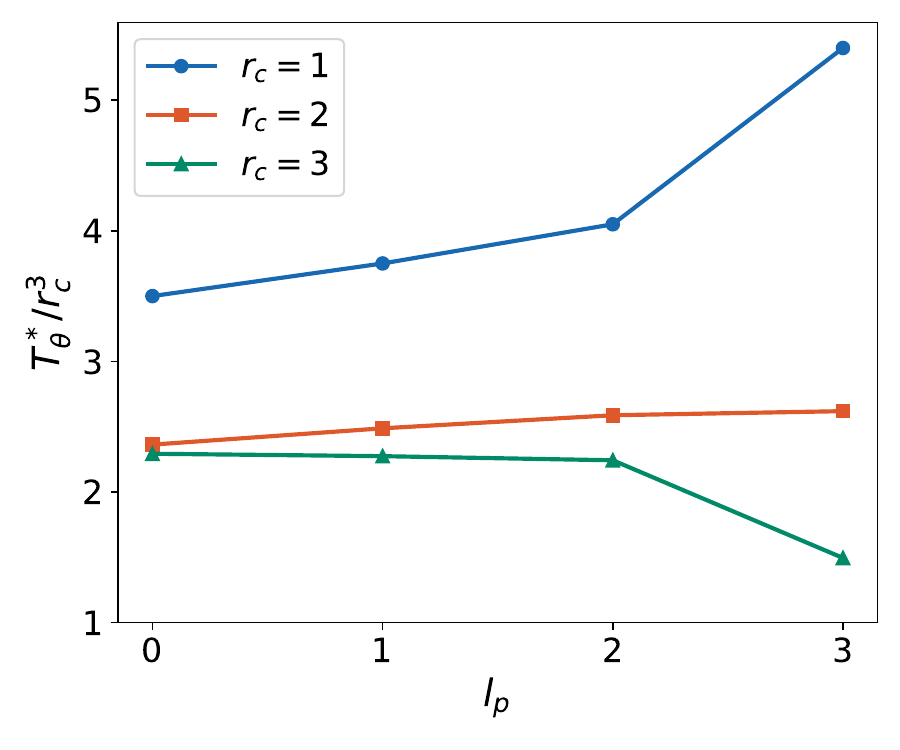}}
\caption{ \textsf{Dependence of $T^{*}_\theta/r_c^3$ on $l_p$ for
    $r_c\,{=}\,1, 2,$ and $3$. All error bars are smaller than the symbol size. The
     chain length in all cases is $N\,{=}\,500$, and the bond length $b$ is taken as the unit length.
     }}
\label{fig4}
\end{figure}

It is instructive to look at
the transition temperature divided by $r_c^3$, 
the volume of the attractive interaction range.
A larger $r_c$ means that more monomers are included within the
interaction range, therefore shifting $T^{*}_\theta$ to higher
values. Looking at $T^*_\theta/r_c^3$ allows for a meaningful
comparison across different $r_c$ values and enables all
$T^{*}_\theta$ values to be presented on the same plot.
Figure~\ref{fig4} shows the dependence of $T^{*}_\theta/r_c^3$ on
$l_p$, for three values of $r_c$. For small $r_c$, $T^{*}_\theta$
increases with the persistence length $l_p$ (blue curve), while for
large $r_c$, $T^{*}_\theta$ decreases with $l_p$ (green curve).

Thus, in addition to influencing the transition sharpness, the interplay between
$l_p$ and $r_c$ also affects the behavior of $T_\theta$. As mentioned
in Sec.~1, there have been conflicting reports 
in the literature ~\cite{Seaton13,Marenz16,Majumder21,Doniach96,Bastolla97} as to whether
$T_\theta$ increases, decreases, or remains unchanged, with increasing
stiffness. Our results and analysis suggest that {\it all three scenarios} are possible.
As can be seen in
Fig.~\ref{fig4}, we find that for small $r_c$, $T_\theta$ increases
with $l_p$, while for large $r_c$, it decreases with $l_p$.
Namely, the manner in which stiffness changes $T_\theta$ depends qualitatively 
on $r_c$, the range of the
attractive pair-potential between monomers.

\begin{figure*}[h!t]
{\includegraphics[width=0.8\textwidth,draft=false]{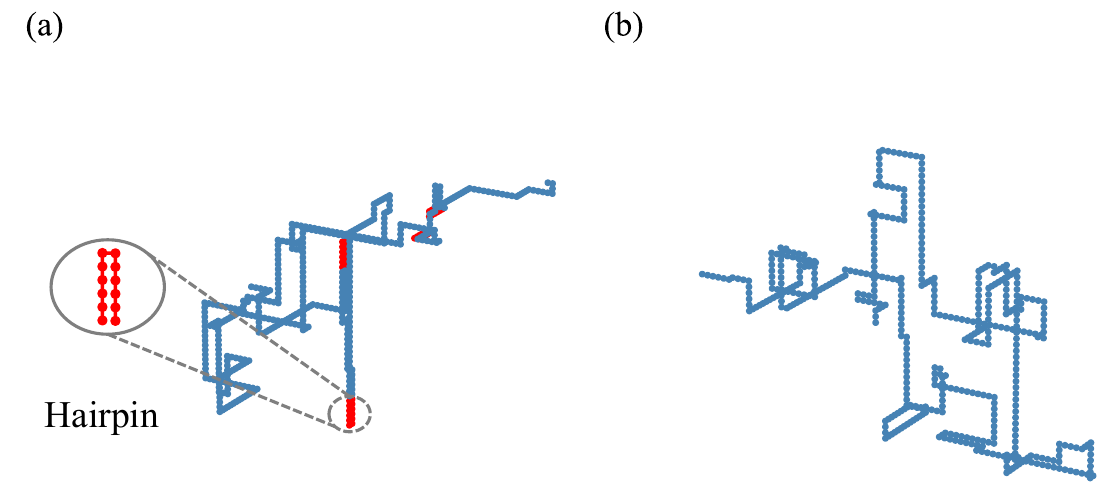}}
\caption{ \textsf{(a) A typical chain configuration for  $l_p\,{=}\,3$ and $r_c\,{=}\,1$ at $T^{*}_\theta=5.4$, 
with the detected hairpin segments highlighted in red. A blow-up of the hairpin segment is also shown.
(b) A typical chain configuration for $l_p\,{=}\,3$ and $r_c\,{=}\,3$ at $T^{*}_\theta=40.3$ with no hairpins.   
In both cases, the chain length is $N\,{=}\,500$.
    }}
\label{fig5}
\end{figure*}

This intriguing behavior can be explained by different mechanisms of contact formation. 
For small $r_c$, a significant attraction requires monomers to come very close together. 
The collapse is then determined by the competition between local bending and short-range attraction. Increasing stiffness favors the formation of local folded structures, such as 
hairpins, and promotes collapse, thereby increasing $T_\theta$ with $l_p$. Conversely, for large $r_c$, 
attractive interactions can occur at greater monomer separations, and the collapse will be less dependent on tight local folds. 
Increasing chain stiffness primarily suppresses the deformations needed for global compaction. Therefore, a stronger attraction (lower temperatures) is necessary, 
resulting in a decrease in $T_\theta$ with $l_p$. These mechanisms, which to our knowledge were not recognized before, demonstrate that the attraction range 
provides a unique physical pathway for collapse that stiffness alone cannot explain.

To illustrate this mechanism, Fig.~\ref{fig5} shows two typical chain configurations obtained from the simulations, one at 
$T^{*}_\theta=5.4$ for $l_p\,{=}\,3$ and $r_c\,{=}\,1$, and the other at $T^{*}_\theta=40.3$ for $l_p\,{=}\,3$ and $r_c\,{=}\,3$. 
A “typical configuration” refers to chains whose $R_g$ is closest to the peak of the $R_g$ distribution. 
Hairpins are clearly observed for $r_c\,{=}\,1$ (Fig.~\ref{fig5}a), but not for $r_c\,{=}\,3$ (Fig.~\ref{fig5}b).
A hairpin is identified when two nonadjacent segments, separated by a contour distance $s$, 
are arranged in an anti-parallel fashion and remain spatially close over a finite segment of the contour. 
Once formed, such hairpins promote chain collapse~\cite{Bastolla97}.

\subsection{Effect of chain length}

\begin{figure}[h!t]
{\includegraphics[width=0.45\textwidth,draft=false]{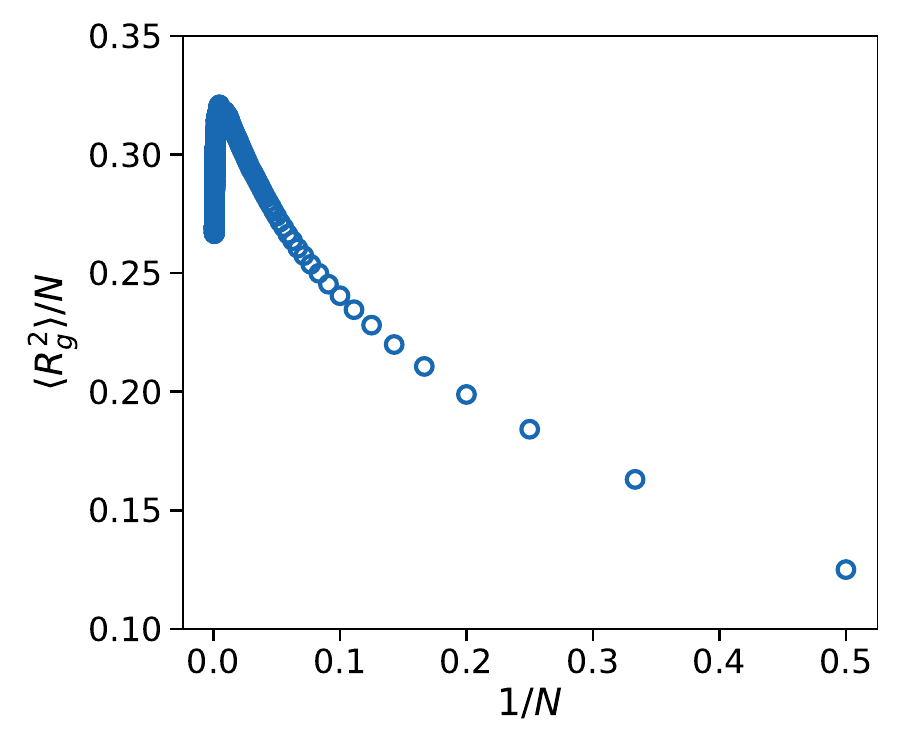}}
\caption{ \textsf{Swelling factor $\langle
      R_g^2\rangle/N$ as a function of $1/N$ at
      fixed temperature $T^*\,{=}\,k_{\rm B} T/\varepsilon\,{=}\,60$, for polymer chains
      of length up to $N\,{=}\,1{,}500$, in the case $l_p\,{=}\,0$ and $r_c\,{=}\,3$.
      }}
\label{fig6}
\end{figure}
  
\begin{figure*}[h!t]
{\includegraphics[width=0.8\textwidth,draft=false]{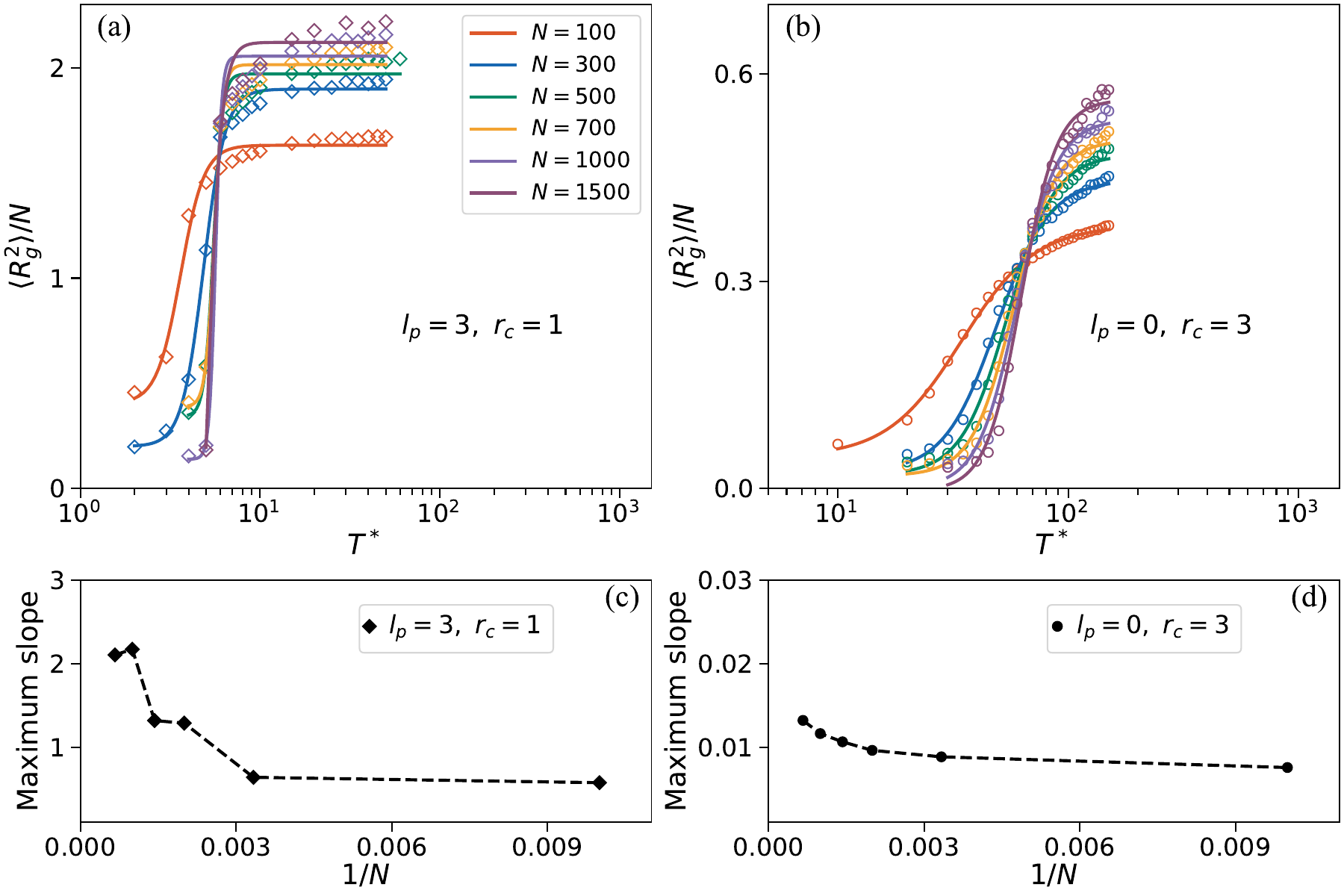}}
\caption{ \textsf{Temperature dependence of $\langle R_g^2\rangle/N$ for chain lengths
$N\,{=}\,100,\, 300,\, 500,\, 700,\, 1{,}000$ and $1{,}500$.
Symbols show simulation data, and solid lines are fits to a sigmoid-like function
[Eq.~(\ref{fit})].
(a) $l_p\,{=}\,3$, $r_c\,{=}\,1$.
(b) $l_p\,{=}\,0$, $r_c\,{=}\,3$.
(c) and (d) Maximum slope of the fitted curves as a function of $1/N$.
The vertical scale in (d) is two orders of magnitude smaller than in (c).
Lines in (c) and (d) are guides to the eye.
Diamonds in (a) and (c) correspond to $l_p\,{=}\,3$, $r_c\,{=}\,1$, and circles in (b) and (d)  to $l_p\,{=}\,0$, $r_c\,{=}\,3$.
      }}
\label{fig7}
\end{figure*}

Having shown how $l_p$ and $r_c$ influence the collapse
transition, we now examine the chain length's role on
the transition temperature and sharpness. 
The main motivation is to analyze the behavior for exceedingly long chains. 
In the preceding sections, we fixed the chain length at $N\,{=}\,500$. However, to examine  
the finite-size effect, we vary the chain length $N$. 
As shown in Fig.~\ref{fig6}, for $l_p\,{=}\,0$ and $r_c\,{=}\,3$, chains of length $N\,{=}\,1{,}500$ ($1/N\,{=}\,6.7 \times 10^{-4}$)
are already collapsed at $T^*\,{=}\,60$, whereas
the shorter chains are not. This observation suggests that the collapse transition
temperature depends on the chain length. As the chain length increases,
the transition temperature is expected to approach the $\theta$-point
for flexible chains in the thermodynamic limit ($N \to \infty$).

The chain length also affects the sharpness of the transition, making it sharper for longer
chains, as is illustrated in Fig.~\ref{fig7}a and~\ref{fig7}b.  Comparing the results for $l_p\,{=}\,3$ and $r_c\,{=}\,1$ 
(Fig.~\ref{fig7}a) with those for $l_p\,{=}\,0$ and $r_c\,{=}\,3$ (Fig.~\ref{fig7}b), we
find that for flexible chains with longer-range attraction ($l_p\,{=}\,0$, $r_c\,{=}\,3$), the slope does not increase significantly but remains
relatively small. Indeed, the curves in Fig.~\ref{fig7}b seem to converge for large $N$ to a curve of finite slope. This
suggests that for $l_p<r_c$, the transition may not
become truly sharp even in the  $N \to \infty$ limit. 

To quantify this behavior, we fit the
$\langle R_g^2\rangle/N$ versus $T^*$ data using a sigmoid-like function [see
 Eq.~(\ref{fit})]. The solid lines in
Fig.~\ref{fig7}a and~\ref{fig7}b show the fitted curves. From these fits, 
we find that the maximum slope of each fitted curve decreases with $1/N$, 
as expected (Fig.~\ref{fig7}c and ~\ref{fig7}d). The value for $l_p\,{=}\,3$,
$r_c\,{=}\,1$ is approximately $100$ times larger than that for $l_p\,{=}\,0$,
$r_c\,{=}\,3$, and it increases much more rapidly with chain length.

Thus, both the sharpness and the temperature of the collapse transition
vary systematically with chain length, although this dependence is
weaker than the influence of $l_p$ and $r_c$.  

To place our results in the context of existing scaling theories, we consider a scaling argument that predicts both shift and rounding of the collapse transition due to finite chain length and chain stiffness~\cite{Wang17}. The deviation of the collapse temperature $T_{\mathrm{tr}}$ 
from the theta point is defined as $\tau_{\mathrm{tr}} = 1-\frac{T_\theta}{T_{\mathrm{tr}}} \approx \frac{T_{\mathrm{tr}}}{T_\theta}-1$, where $T_\theta$ 
is the collapse temperature for flexible and infinite chains. Then, $\tau_{\mathrm{tr}}$ scales as
\be
|\tau_{\mathrm{tr}}| \simeq N^{-1/2} \left( \frac{b^3}{v_{\mathrm{m}}} \right)^{1/2},
\ee
and the width of the transition is 
\be
|\delta \tau_{\mathrm{tr}}| \simeq N^{-1/2} \left( \frac{b^3}{v_{\mathrm{m}}} \right)^{-1/2},
\ee
reflecting the combined effects of the finite chain length $N$, monomer size $b$, and an effective excluded-volume parameter $v_{\mathrm{m}}$, related to chain stiffness. 
One might expect that the attraction range would modify the effective two-body interaction, much as stiffness modifies the excluded volume. However, our results show 
that the role of $r_c$ cannot be reduced to just renormalizing the excluded-volume parameter or stiffness. In fact, the broadening of the transition in the scaling expression vanishes as $N$ approaches infinity, whereas a large $r_c$ causes broadening for arbitrarily long chains (Fig.~\ref{fig7}).

We have examined whether the data could be collapsed using a single dimensionless variable such as $l_p/r_c$, 
but we find that a complete data collapse is not achieved. In particular, different values of $r_c$ lead to qualitatively different trends of $T_\theta$ 
as a function of $l_p$ for fixed $l_p/r_c$, indicating that the collapse behavior cannot be described by a simple single-parameter scaling. Instead, both stiffness 
and attraction range must be treated as independent control parameters. Alternatively, a dependence on at least two dimensionless length ratios, such as $l_p/L$ and $l_p/r_c$, 
should be considered (assuming that the monomer size $b$ can be taken to zero). A more complete theoretical description would need to incorporate these two effects explicitly. 
However, these additional considerations lie outside the scope of the present work.

Increasing $r_c$ cannot be seen solely as a renormalization of the persistence length, for two reasons. First, it qualitatively changes how $T_\theta$ depends on $l_p$. Specifically, the role of stiffness is not fixed but varies with $r_c$ and can even change sign across different regimes (Fig.~\ref{fig4}). Second, a large $r_c$ 
may eliminate the critical collapse transition, transforming it into a gradual contraction over a finite temperature range, even for an infinitely long chain (Fig.~\ref{fig7}). 
This suggests that the attraction range introduces an additional physical mechanism beyond simple stiffness rescaling, resulting in fundamentally different collapse behavior.

\section{Conclusions}

\begin{figure*}[h!t]
{\includegraphics[width=0.6\textwidth,draft=false]{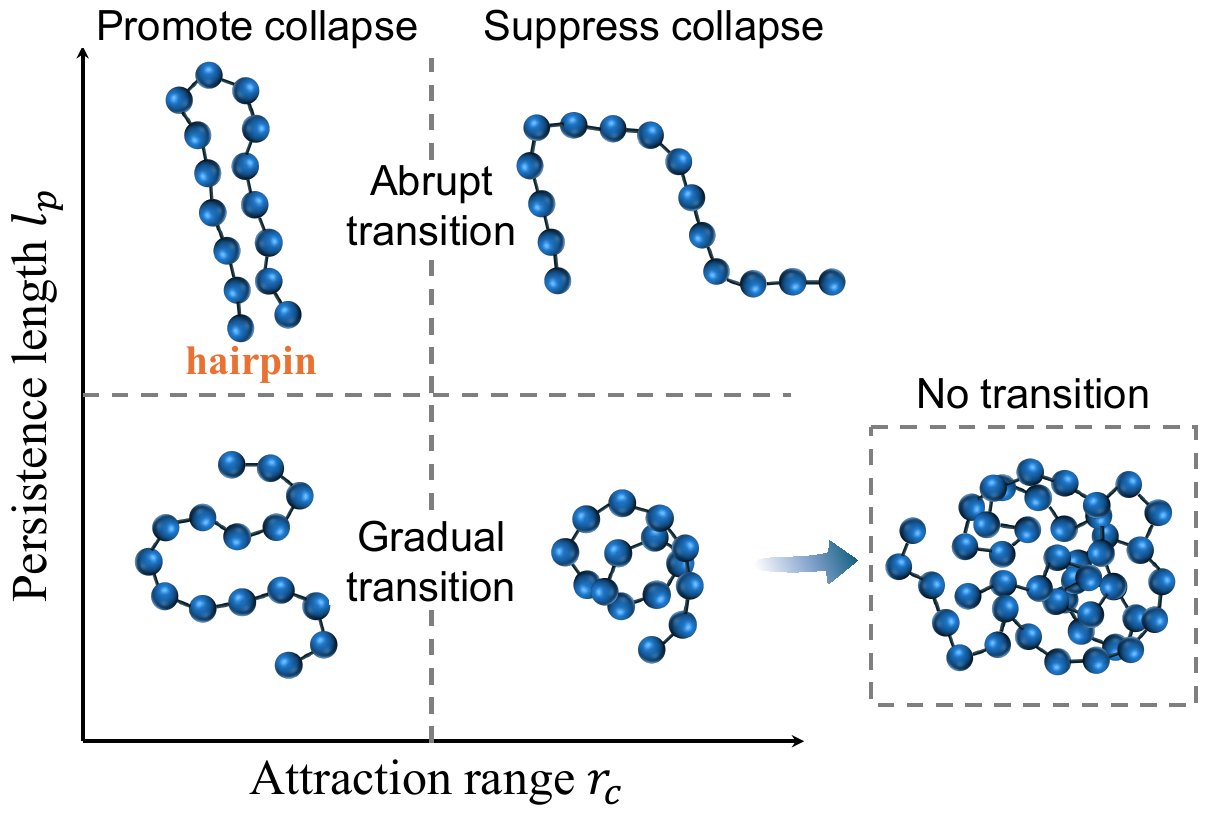}}
\caption{ \textsf{A schematic illustration of the collapse behavior of a single polymer chain,
 summarizing the roles of the persistence length $l_p$ and the attraction range $r_c$.
}}
\label{fig8}
\end{figure*}

In this study, we have investigated the single-chain collapse of
polymer chains using Monte Carlo simulations based on the
Pruned-Enriched Rosenbluth Method (PERM). We have analyzed the
influence of two key parameters: the persistence length, $l_p$,
associated with the intrinsic stiffness of the polymer chain, and
$r_c$, determining the range of monomer-monomer attraction. 
We propose a physical model in which the competition between $l_p$ and $r_c$ determines the chain-collapse behavior. We also offer a physical 
explanation for recent experimental results on ssRNA and dsDNA, as reported in Ref.~\citenum{Knobler23}. A key result is that a large attraction range, $r_c$, 
can turn the critical collapse transition into a gradual crossover, and this finite-width crossover remains as the chain length grows. Additionally, 
we show that the way the transition temperature $T_\theta$ depends on $l_p$ is qualitatively altered by $r_c$. 
Figure~\ref{fig8} summarizes our main findings in the $l_p$-$r_c$ 
plane, illustrating how the competition between chain stiffness and attraction range determines 
both the sharpness of the collapse and whether stiffness promotes or suppresses it.

Figure~\ref{fig8} shows that increasing the persistence length leads to 
a sharper collapse transition, whereas expanding the attraction range
results in a more gradual collapse. Overall, our results indicate that the sharpness of 
the collapse transition depends on the competition between stiffness and interaction range.  When $l_p\gtrsim r_c$, 
stiff chains undergo a sharp and well-defined collapse transition. By contrast, when $l_p\lesssim r_c$, the critical collapse transition 
is effectively removed, turning into gradual compaction over a finite temperature range. Figure~\ref{fig7} further suggests
that this rounding persists upon increasing the chain length.
In cases where the attraction is induced by condensing 
molecules, this removal of criticality is supported by a scaling argument (see Supporting Information).

 We have examined the dependence of the transition temperature
$T_\theta$ on chain stiffness. We have found that stiffness may either
increase $T_\theta$ or decrease it, depending on the attraction
range $r_c$. For small $r_c$, increasing stiffness promotes collapse
and raises $T_\theta$, whereas for large $r_c$, it suppresses collapse
and lowers $T_\theta$. These results (see Fig.~\ref{fig4}) provide important insight into
the relative role of polymer stiffness and monomer interactions in
relation to the collapse transition. This new insight may resolve
previous conflicting results concerning the effect of stiffness on the
transition temperature.

The polymer chain length $N$ plays a secondary yet
systematic role. 
Longer chains exhibit a sharper transition and a slightly shifted transition temperature. However, this effect is
significantly weaker than that of $l_p$ and $r_c$. This suggests
that, while stiffness and interaction range dominate the collapse
behavior, finite-size effects must be considered when interpreting
simulation or experimental data.

Our findings contribute to a more systematic understanding of the polymer phase
behavior and may have valuable implications for designing responsive
polymeric materials, biomolecular folding, and nanotechnology
applications~\cite{Sarker10,Patel17,Artem24}. The main conclusion is that a sharp, 
sensitively controlled collapse (e.g., as in DNA condensation \cite{Knobler23}) requires chain stiffness,
and a more specific criterion is $l_p\gtrsim r_c$.  Future research can
incorporate solvent effects, adding condensing agents such as
multivalent ions, as in the experiments of ref~\citenum{Knobler23}, or surfactants, and exploring more complex
architectures beyond linear polymer chains.

\section*{Supporting Information}
Scaling Argument for the Rounding of the Collapse Transition.

\section*{Acknowledgments}
We are grateful to A.\ Giacometti for particularly insightful and helpful discussions; we also thank H.\ Orland and R.\ Zandi for useful discussions. This work was partly supported by the Israel Science Foundation (ISF) under Grants Nos. 226/24 and 1611/24.

\newpage


\begin{thebibliography}{99}

\bibitem{Grosberg94} Grosberg, A. Y.; Khokhlov, A. R. \emph{Statistical Physics of Macromolecules}; AIP Press, New York, \textbf{1994}.

\bibitem{Rubinstein03} Rubinstein, M.; Colby, R. H. \emph{Polymer Physics}; Oxford University Press, New York, \textbf{2003}.

\bibitem{Kita05} Kita, R.; Wiegand, S. Soret Coefficient of Poly(N-isopropylacrylamide)/Water in the Vicinity of Coil-Globule Transition Temperature. \emph{Macromolecules} \textbf{2005}, \emph{38}, 
4554-4556.

\bibitem{Frerix06} Frerix, A.; Schonewald, M.; Geilenkirchen, P.; M\"uller, M.; Kula, M. R.; Hubbuch, J. 
Exploitation of the Coil-Globule Plasmid DNA Transition Induced by Small Changes in Temperature, pH, Salt, and Poly(ethylene glycol) 
Compositions for Directed Partitioning in Aqueous Two-Phase Systems. \emph{Langmuir} \textbf{2006}, \emph{22}, 4282-4290. 

\bibitem{Tanaka09} Tanaka, F.; Koga, T.; Kojima, H.; Winnik, F. M. Temperature- and Tension-Induced Coil-Globule Transition of Poly(N-isopropylacrylamide) 
Chains in Water and Mixed Solvent of Water/Methanol. \emph{Macromolecules} \textbf{2009}, \emph{42}, 1321-1330.

\bibitem{Nakata95} Nakata, M. Coil-Globule Transition of Poly(methyl methacrylate) in a Mixed Solvent. \emph{Phys. Rev. E} \textbf{1995}, \emph{51}, 5770.

\bibitem{Sherman06} Sherman, E.; Haran, G. Coil-Globule Transition in the Denatured State of a Small Protein. \emph{Proc. Natl. Acad. Sci. U.S.A.} \textbf{2006}, \emph{103}, 11539-11543.

\bibitem{Xu06} Xu, J.; Zhu, Z.; Luo, S.; Wu, C.; Liu, S. First Observation of Two-Stage Collapsing Kinetics of a Single Synthetic Polymer Chain. \emph{Phys. Rev. Lett.} \textbf{2006}, \emph{96}, 027802.

\bibitem{Stockmayer60} Stockmayer, W. H. Problems of the Statistical Thermodynamics of Dilute Polymer Solutions. \emph{Makromol. Chem.} \textbf{1960}, \emph{35}, 54-74.

\bibitem{Tanaka79} Nishio, I.; Sun, S. T.; Swislow, G.; Tanaka, T. First Observation of the Coil-Globule Transition in a Single Polymer Chain. \emph{Nature} \textbf{1979}, \emph{281}, 208-209.

\bibitem{Tanaka80} Swislow, G.; Sun, S. T.; Nishio, I.; Tanaka, T. Coil-Globule Phase Transition in a Single Polystyrene Chain in Cyclohexane. \emph{Phys. Rev. Lett.} \textbf{1980}, \emph{44}, 796.

\bibitem{Lifshitz68} Lifshitz, I. M. Some Problems of the Statistical Theory of Biopolymers. \emph{Zh. Eksp. Teor. Fiz.} \textbf{1968}, \emph{55}, 2408-2422. (Engl. Transl.: Sov. Phys. JETP 
\textbf{1969}, \emph{28}, 1280-1286).

\bibitem{Lifshitz78} Lifshitz, I. M.; Grosberg, A. Y.; Khokhlov, A. R. Some Problems of the Statistical Physics of Polymer Chains with Volume Interaction. \emph{Rev. Mod. Phys.} \textbf{1978}, \emph{50}, 683-713.

\bibitem{Grosberg92} Grosberg, A. Y.; Kuznetsov, D. V. Quantitative Theory of the Globule-to-Coil Transition. 1. Link Density Distribution in a Globule and Its Radius of Gyration. 
\emph{Macromolecules} \textbf{1992}, \emph{25}, 1970-1979.

\bibitem{Grosberg922} Grosberg, A. Y.; Kuznetsov, D. V. Quantitative Theory of the Globule-to-Coil Transition. 4. Comparison of Theoretical Results with Experimental Data. 
\emph{Macromolecules} \textbf{1992}, \emph{25}, 1996-2003.

\bibitem{Wu98} Wu, C.; Wang, X. Globule-to-Coil Transition of a Single Homopolymer Chain in Solution. \emph{Phys. Rev. Lett.} \textbf{1998}, \emph{80}, 4092-4094.

\bibitem{Wu96} Wu, C.; Zhou, S. First Observation of the Molten Globule State of a Single Homopolymer Chain. \emph{Phys. Rev. Lett.} \textbf{1996}, \emph{77}, 3053.

\bibitem{Baysal98} Baysal, B. M.; Kayaman, N. Coil-Globule Transition of Poly(methyl methacrylate) by Intrinsic Viscosity. \emph{J. Chem. Phys.} \textbf{1998}, \emph{109}, 8701.

\bibitem{Ye07} Ye, X.; Lu, Y.; Shen, L.; Ding, Y.; Liu, S.; Zhang, G.; Wu, C. How Many Stages Are in the Coil-to-Globule Transition of Linear Homopolymer Chains in a Dilute Solution? 
\emph{Macromolecules} \textbf{2007}, \emph{40}, 4750.

\bibitem{Chakraborty18} Chakraborty, I.; Mukherjee, K.; De, P.; Bhattacharyya, R. Monitoring Coil-Globule Transitions of Thermoresponsive Polymers by Using NMR Solvent Relaxation. 
\emph{J. Phys. Chem. B} \textbf{2018}, \emph{122}, 6094.

\bibitem{Bustamante94} Bustamante, C.; Marko, J. F.; Siggia, E. D.; Smith, S. Entropic Elasticity of $\lambda$-Phage DNA. \emph{Science} \textbf{1994}, \emph{265}, 1599-1600.

\bibitem{Gerrits21} Gerrits, L.; Hammink, R.; Kouwer, P. H. J. Semiflexible Polymer Scaffolds: An Overview of Conjugation Strategies. \emph{Polym. Chem.} \textbf{2021}, \emph{12}, 1362-1392.

\bibitem{Tatjana19} Skrbic, T.; Banavar, J. R.; Giacometti, A. Chain Stiffness Bridges Conventional Polymer and Biomolecular Phases. \emph{J. Chem. Phys.} \textbf{2019}, \emph{151}, 174901.

\bibitem{Arcangeli24} Arcangeli, T.; Skrbic, T.; Azote, S.; Marcato, D.; Rosa, A.; Banavar, J. R.; Piazza, R.; Maritan, A.; Giacometti, A. Phase Behavior and Self-Assembly of 
Semiflexible Polymers in Poor-Solvent Solutions. \emph{Macromolecules} \textbf{2024}, \emph{57}, 8940-8955.

\bibitem{Leforestier09} Leforestier, A.; Livolant, F. Structure of Toroidal DNA Collapsed Inside the Phage Capsid. \emph{Proc. Natl. Acad. Sci. U.S.A.} \textbf{2009}, \emph{106}, 9157-9162.

\bibitem{Leforestier11} Leforestier, A.; Siber, A.; Livolant, F.; Podgornik, R. Protein-DNA Interactions Determine the Shapes of DNA Toroids Condensed in Virus Capsids. \emph{Biophys. J.} 
\textbf{2011}, \emph{100}, 2209-2216.

\bibitem{Seaton13} Seaton, D. T.; Schnabel, S.; Landau, D. P.; Bachmann, M. From Flexible to Stiff: Systematic Analysis of Structural 
Phases for Single Semiflexible Polymers. \emph{Phys. Rev. Lett.} \textbf{2013}, \emph{110}, 028103.

\bibitem{Marenz16} Marenz, M.; Janke, W. Knots as a Topological Order Parameter for Semiflexible Polymers. \emph{Phys. Rev. Lett.} \textbf{2016}, \emph{116}, 128301.

\bibitem{Majumder21} Majumder, S.; Marenz, M.; Paul, S.; Janke, W. Knots Are Generic Stable Phases in Semiflexible Polymers. \emph{Macromolecules} \textbf{2021}, \emph{54}, 5321-5334.

\bibitem{Yang13} Yang, D.; Wang, Q. Unified View on the Mean-Field Order of Coil-Globule Transition. \emph{ACS Macro Lett.} \textbf{2013}, \emph{2}, 952-954.

\bibitem{Wang17} Wang, Z. G. 50th Anniversary Perspective: Polymer Conformation -- A Pedagogical Review. \emph{Macromolecules} \textbf{2017}, \emph{50}, 9073-9114.

\bibitem{Post79} Post, C. B.; Zimm, B. H. Internal Condensation of a Single DNA Molecule. \emph{Biopolymers} \textbf{1979}, \emph{18}, 1487-1501.

\bibitem{Rampf06} Rampf, F.; Binder, K.; Paul, W. The Single Polymer Chain Phase Diagram: New Insights from a New Simulation Method. 
\emph{J. Polym. Sci., Part B: Polym. Phys.} \textbf{2006}, \emph{44}, 2542-2555.

\bibitem{Noguchi97} Noguchi, H.; Yoshikawa, K. First-Order Phase Transition in a Stiff Polymer Chain. \emph{Chem. Phys. Lett.} \textbf{1997}, \emph{278}, 184-188.

\bibitem{Doniach96} Doniach, S.; Garel, T.; Orland, H. Phase Diagram of a Semiflexible Polymer Chain in a $\theta$ Solvent: Application to Protein Folding. 
\emph{J. Chem. Phys.} \textbf{1996}, \emph{105}, 1601.

\bibitem{Bastolla97} Bastolla, U.; Grassberger, P. Phase Transitions of Single Semistiff Polymer Chains. \emph{J. Stat. Phys.} \textbf{1997}, \emph{89}, 1061-1078.

\bibitem{Binder08} Binder, K.; Paul, W.; Strauch, T.; Rampf, F.; Ivanov, V.; Luettmer-Strathmann, J. Phase Transitions of Single Polymer Chains and of Polymer Solutions: 
Insights from Monte Carlo Simulations. \emph{J. Phys.: Condens. Matter} \textbf{2008}, \emph{20}, 494215.

\bibitem{Binder09} Taylor, M. P.; Paul, W.; Binder, K. Phase Transitions of a Single Polymer Chain: A Wang-Landau Simulation Study. 
\emph{J. Chem. Phys.} \textbf{2009}, \emph{131}, 114907.

\bibitem{Binder13} Taylor, M. P.; Paul, W.; Binder, K. Applications of the Wang-Landau Algorithm to Phase Transitions of a Single Polymer Chain. \emph{Polymer Sci. Ser. C} \textbf{2013}, \emph{55}, 23-38.

\bibitem{Knobler23} Duran-Meza, A. L.; Oster, L.; Sportsman, R.; Phillips, M.; Knobler, C. M.; Gelbart, W. M. Long ssRNA Undergoes Continuous Compaction in the 
Presence of Polyvalent Cations. \emph{Biophys. J.} \textbf{2023}, \emph{122}, 3469-3475.

\bibitem{Haim01} Diamant, H.; Andelman, D. General Criterion for Controllable Conformational Transitions of Single- and Double-Stranded DNA. arXiv, September 13, 2001. https://doi.org/10.48550/arXiv.cond-mat/0109253 (accessed 2026-03-18).

\bibitem{Grassberger97} Grassberger, P. Pruned-Enriched Rosenbluth Method: Simulations of $\theta$ Polymers of Chain Length up to 1~000~000. 
\emph{Phys. Rev. E} \textbf{1997}, \emph{56}, 3682.

\bibitem{Rosenbluth55} Rosenbluth, M. N.; Rosenbluth, A. W. Monte Carlo Calculation of the Average Extension of Molecular Chains. \emph{J. Chem. Phys.} \textbf{1955}, \emph{23}, 356.

\bibitem{Wall59} Wall, F. T.; Erpenbeck, J. J. New Method for the Statistical Computation of Polymer Dimensions. \emph{J. Chem. Phys.} \textbf{1959}, \emph{30}, 634.

\bibitem{Hsu11} Hsu, H. P.; Grassberger, P. A Review of Monte Carlo Simulations of Polymers with PERM. \emph{J. Stat. Phys.} \textbf{2011}, \emph{144}, 597-637.

\bibitem{Golestanian99} Golestanian, R.; Kardar, M.; Liverpool, T. B. Collapse of Stiff Polyelectrolytes due to Counterion Fluctuations. \emph{Phys. Rev. Lett.} \textbf{1999}, \emph{82}, 4456.

\bibitem{Bloomfield97} Bloomfield, V. A. DNA Condensation by Multivalent Cations. \emph{Biopolymers} \textbf{1997}, \emph{44}, 269–282.

\bibitem{Gelbart00} Gelbart, W. M.; Bruinsma, R. F.; Pincus, P. A.; Parsegian, V. A. DNA-Inspired Electrostatics. \emph{Phys. Today} \textbf{2000}, \emph{53}, 38–44.

\bibitem{Castelnovo04} Castelnovo, M.; Gelbart, W. M. Semiflexible Chain Condensation by Neutral Depleting Agents: Role of Correlations between Depletants. \emph{Macromolecules} \textbf{2004}, \emph{37}, 3510-3517.

\bibitem{Sarker10} Shepherd, J.; Sarker, P.; Swindells, K.; Douglas, I.; MacNeil, S.; Swanson, L.; Rimmer, S. 
Binding Bacteria to Highly Branched Poly(N-Isopropyl Acrylamide) Modified with Vancomycin Induces the Coil-to-Globule Transition. \emph{J. Am. Chem. Soc.} \textbf{2010}, \emph{132}, 1736-1737.

\bibitem{Patel17} Patel, A.; Malinovska, L.; Saha, S.; Wang, J.; Alberti, S.; Krishnan, Y.; Hyman, A. A. ATP as a Biological Hydrotrope. \emph{Science} \textbf{2017}, \emph{356}, 753-756.

\bibitem{Artem24} Rumyantsev, A. M.; Gavrilov, A. A.; Johner, A. Complete Diagram of Conformational Regimes for Polyampholytic Disordered Proteins. 
\emph{Macromolecules} \textbf{2024}, \emph{57}, 5533-5544.


\end{thebibliography}
\end{document}



\section*{Supporting Information}

\section {Scaling argument}
We present a heuristic scaling argument for the competing effects of the range of attraction,
 induced by condensing agents, and the persistence length, on the collapse transition
\cite{Haim01}.  The analysis is valid in the limit $b \ll l_p,r_c \ll
Nb$. Hence, it cannot be directly related to the numerical results
presented in the main text.
The purpose of the scaling argument is to rationalize the
trends we observed in the numerical simulations in a different context. Specifically, we explore the conditions under which
an ordinary coil-to-globule transition, occurring at a critical
temperature, can be turned by condensing agents into a gradual contraction occurring over a finite
temperature range in the thermodynamic limit, $N\rightarrow\infty$.

We express the attraction between any two monomers as
\be
\label{u_att}
u_{\rm att}(r)=-Af(r/r_c) \,,
\ee
where $r$ is the distance between the monomers in 3D space, $A$ is the
attraction strength, assumed to be small compared to the thermal
energy $k_{\rm B}T$, $r_c$ is the attraction range, and $f(r/r_c)$ is a
function that decays fast to zero for $r > r_c$. 
In various scenarios, such as the one considered in
Ref.~\citenum{Knobler23}, the attraction between monomers does not come
from a direct potential but is induced by condensing agents. In such a
case $A$ and $r_c$ would not be constant parameters but depend on
external conditions, such as the agent concentration. 
Consequently, in our scaling argument, we treat $A$ and $r_c$
as parameters that can be tuned by the condensing agents.

The attraction reduces the excluded-volume parameter of the
chain, $v(T)\,{=}\,v_0-\delta v(T)$, where $v_0$ is the excluded-volume
parameter in the absence of the attraction, and
\begin{eqnarray}
\delta v(T) &=& -\int {\rm d}^3 {r} \left[ 1 - e^{-\beta u_{\rm att}(r)} \right] \simeq
-\beta\int {\rm d}^3 {r}\,  u_{\rm att}(r) \nonumber\\
&\sim& \beta A r_c^3 ,
 \label{eq:delta}
\end{eqnarray}
recalling that $\beta\,{=}\,1/k_{\rm B} T$.

When the overall monomer-monomer interaction (consisting of
self-avoidance and attraction) is repulsive ($v(T)>0$), the chain is
swollen, whereas when it is attractive ($v(T)<0$), the chain is
collapsed~\cite{Gennes79}. In the usual situation, where the competing
repulsive and attractive interactions have similar (short) ranges, the
collapse transition occurs at the sharply defined temperature
$T_\theta$ for which $v(T_\theta)\,{=}\,0$~\cite{Gennes79}.

In a different scenario of gradual contraction, the collapse does not
involve all the monomers. We show that the dominance of different interactions on 
different length scales can bring this about. If the
attraction is weak but long-ranged, it will dominate over large
distances, while at smaller distances, the repulsive self-avoidance
will be the dominant interaction. In such a scenario, the chain can be
envisioned as divided into subunits, or `blobs' (a useful concept
introduced by de Gennes and often used in polymer
physics~\cite{Gennes79}).

\begin{figure}[h!t]
{\includegraphics[width=0.25\textwidth,draft=false]{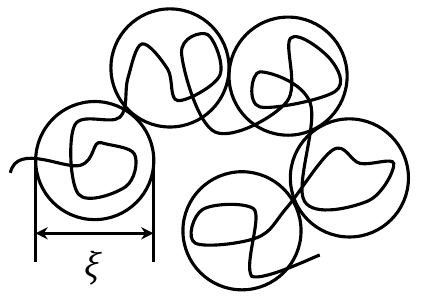}}
\caption{ \textsf{A schematic illustration of a chain of blobs with blob size $\xi$.
}}
\label{fig9}
\end{figure}

Imagine that the chain is divided into blobs of spatial size $\xi$,
each containing $\textsl{g}$ monomers, as illustrated schematically in Fig.~\ref{fig9}. The behavior of the rescaled
`chain of blobs', in which each blob is an effective monomer, is
determined by an {\it effective interaction} $U_{\rm eff}(r)$ acting between two
blobs. This consists of strong inter-blob repulsion at distances
smaller than $\xi$, and attraction at distances larger than $\xi$ due
to the integrated attractive interaction of $\textsl{g}^2$ monomer pairs 
in the two blobs,
\be
\label{U_blob}
U_{\rm eff}(r) =
\begin{cases}
\infty~, &\quad r< \xi \\
\textsl{g}^2u_{\rm att}(r) ~, &\quad r > \xi .
\end{cases}
\ee
This effective interaction implies that the rescaled chain of blobs is characterized by 
an effective excluded-volume parameter, $V_{\rm eff}$
\be
\label{V}
V_{\rm eff}(T) = \int {\rm d}^3 r \left[1-{\rm e}^{-\beta U_{\rm eff}(r) }\right].
\ee
Although we have assumed weak monomer-monomer attraction, $A\,{\ll}\, k_{\rm B}T$ [Eq.~(\ref{u_att})],
the effective attraction between blobs, proportional to $\textsl{g}^2
A$, is not necessarily smaller than $k_{\rm B}T$. This necessitates the 
accurate expression for the excluded-volume parameter, Eq.~(\ref{V}).

When the chain is in the temperature range of such gradual
contraction, the rescaled chain of blobs is in the critical regime.
As $T$ is changed within this range, the blob parameters
$\xi$ and $\textsl{g}$ adapt to keep the rescaled chain at its collapse
transition. Thus, the balance between the competing interactions is
such that the rescaled excluded-volume parameter vanishes, $V_{\rm eff}(T)\,{=}\,0$.
Substituting the expressions for $U_{\rm eff}(r)$ [Eq.~(\ref{U_blob})] and
$u_{\rm att}(r)$ [Eq.~(\ref{u_att})] in Eq.~(\ref{V}), we can rewrite
the condition $V_{\rm eff}\,{=}\,0$ as
\be
\int_1^{\infty} \rho^2 \,\mathrm{d}\rho
\left[
\exp\!\left(\beta \textsl{g}^2 A f\!\left((\xi/r_c)\rho\right)\right)
- 1
\right]
= \text{const.}
\label{eq:int}
\ee
where $\rho\,{=}\,r/\xi$, and the constant on the right-hand side is of
order unity.

The ability of the chain to adjust its subdivision into blobs requires
that Eq.~(\ref{eq:int}) be satisfied for any temperature within the
gradual-contraction range. This can be achieved only if
$\beta\textsl{g}^2 A$ as well as $\xi/r_c$ remain constant. 
\be
\xi \sim r_c,\ \ \ \ \textsl{g} \sim (\beta A)^{-1/2}.
\label{xig}
\ee
Since we have assumed $\beta A\ll 1$, we get $\textsl{g}\gg 1$, consistently with the blob picture. 

At the same time, the size and number of monomers in a blob, $\xi$ and $\textsl{g}$, are geometrically related.
For $\textsl{g}\gg 1$, they satisfy a scaling relation,
\be
\xi \sim b \textsl{g}^\nu,
\label{nu}
\ee
where $\nu$ is the chain Flory exponent on the blob scale. Using Eqs.~(\ref{xig}) and (\ref{nu}) in Eq.~(\ref{eq:delta}),
we obtain
\be
\textsl{g} \sim (\delta v/b^3)^{1/(3\nu-2)}, \quad \xi \sim (\delta v/b^3)^{\nu/(3\nu-2)}.
\label{eq:cri}
\ee

We now impose a condition of self-consistency on the scenario of
gradual contraction through subdivision. The contraction of the chain
is caused by a decrease of its excluded-volume parameter,
$v\,{=}\,v_0-\delta v$, {\it i.e.,} an increase in $\delta v$. As the chain
contracts, the number of monomers in a blob, $\textsl{g}$, must
decrease. This process stops when $\textsl{g}\sim 1$, resulting in a fully
collapsed chain. According to Eq.~(\ref{eq:cri}), $\textsl{g}$ would
decrease with increasing $\delta v$ only if
\be
\nu < 2/3.
\ee
We recall that the value of $\nu$ is the one applicable within a blob,
{\it i.e.,} corresponding to the chain without the added attraction. If the
chain has a persistence length $l_p$ which is larger than the blob,
then within a blob the chain is rod-like, $\nu\,{=}\,1$, 
and the self-consistency condition is not fulfilled. If,
however, $l_p$ is smaller than the blob, the chain is swollen, $\nu \geq 1/2$, and
gradual contraction is self-consistent. 

The blob size $\xi$ controls the local chain statistics in the scaling argument,
but along the contraction, according to Eq.~(\ref{xig}), the blob size is set by the induced attraction range $r_c$. 
Thus, $\xi$ and $r_c$ are of the same order, and the criterion for sharp versus 
gradual collapse reduces to comparing $l_p$ with $r_c$.
\begin{eqnarray}
l_p\,&\gtrsim&\, r_c: \quad \quad \text {sharp transition} \nonumber \\
l_p\,&\lesssim&\, r_c: \quad\quad \text {gradual contraction} .
 \label{eq:case}
\end{eqnarray}

The scaling argument relies on a mechanism in which the strength and range of the attraction can change
 (e.g., with temperature or concentration of the condensing agents). While this mechanism is not realized in our simulations,
 where the attraction is fixed and strictly vanishes beyond $r_c$, 
the purpose of this analysis is to demonstrate a concrete mechanism by which an increased
range of monomer-monomer attraction is related to the rounding of the critical
collapse transition. This relation is qualitatively in line with
our numerical observation, indicating that the comparison between $l_p$ and $r_c$ controls the sharpness of the transition.

\newpage